\documentstyle[prl,aps,twocolumn,psfig]{revtex}
\begin{document}
\draft
\title{Single-vortex-induced voltage steps in Josephson-junction arrays}
\author{ P. H. E. Tiesinga$^1$,T. J. Hagenaars$^{1,3}$, J. E. van Himbergen$^1$
and Jorge V. Jos\'{e}$^{2}$ \\
{\it $^1$Instituut voor Theoretische Fysica,\\ Princetonplein 5,
Postbus 80006, 3508 TA Utrecht, The Netherlands\\
$^2$Department of Physics and Center for Interdisciplinary
Research on Complex Systems, Northeastern University,\\ Boston
Massachusetts 02115, USA\\
 $^3$ Institut f\"ur Theoretische Physik,
Universit\"at W\"urzburg,\\Am Hubland, 97074 W\"urzburg, Germany }
}

\maketitle
\begin{abstract}
We have numerically and analytically studied ac+dc driven 
Josephson-junction arrays with a single vortex or with a 
single vortex-antivortex pair present.
We find single-vortex steps in the voltage versus 
current characteristics ($I$-$V$) of the array. They 
correspond microscopically to a single vortex
 phase-locked to move a fixed number of plaquettes per period
 of the ac driving current. In underdamped arrays we find vortex
 motion period doubling on the steps. We observe subharmonic steps 
in both underdamped and overdamped arrays.  We successfully 
compare these results with a phenomenological model of vortex motion 
with a nonlinear viscosity. The $I$-$V$ of an array with a 
vortex-antivortex pair displays fractional voltage steps. 
A possible connection of these results to present day experiments
is also discussed.
\end{abstract}
\section{Introduction}
The presence of giant Shapiro steps and giant fractional Shapiro steps
 in the $I$-$V$ characteristics of 2-D  Josephson-junction arrays 
 has attracted significant attention recently 
\cite{GSS1,GSS_VL,GSSUB,GFSIM,GFLAD,ACVS,HERIND}.
 These 2-D arrays may be of use as a source of coherent 
microwave radiation \cite{COH}. In a separate context
 the flux-flow dynamics of vortices has been studied 
\cite{DEPIN,SJMOD,EXPRSS}. The reported experimental 
observation of ballistic vortex motion \cite{BALL}
 has also stimulated further theoretical and numerical 
investigations \cite{V1SIM,Uli,YuR,HYSTSIM,Hag} of
 the mass and friction of a vortex in an array. Until now these numerical 
investigations have focused on dc driven vortices.
In this work we perform numerical 
simulations on Josephson-junction arrays, with
 only one vortex or with a vortex-antivortex pair present in it,
 driven by a time-dependent current 
$i(t)=i_{dc}+i_{ac}\cos(2 \pi \nu t)$. We calculate the voltage
 $V$ versus $i_{dc}$ characteristics ($I$-$V$). 
We find harmonic and subharmonic single-vortex voltage steps 
 and analyze the underlying phase-locked vortex motion.
 A vortex-antivortex pair separated by a distance 
$\Delta x$ along the direction of the injected external current
  phase-locks on to fractional voltage steps.\\
The arrays are 2-D lattices of superconducting islands (sites)
connected by Josephson junctions (bonds). The unit cells 
(plaquettes) of these lattices can be, for example, square or triangular.
The vortices are represented by eddy-current patterns 
about a plaquette. Here we consider the classical regime defined 
by $E_{J}\gg E_{c}=e^2/2C$, where $E_J$ is the Josephson coupling 
energy and $E_c$ the charging energy of two islands, $e$ the electron charge,
and $C$ the capacitance of a junction.
In this regime quantum fluctuations are neglected, leaving 
the phases $\theta (\bbox{r})$ of the Ginzburg-Landau order parameter on the
islands as the only dynamical variables. 

In this case the array is well-modeled by the Resistively Capacitively 
Shunted Junction (RCSJ) model, defined by the total bond
current $i(\bbox{r},\bbox{r'})$ between nearest  neighbor
sites $\bbox{r}$ and $\bbox{r'}$
\begin{eqnarray}
i(\bbox{r},\bbox{r'})&=&
\beta_{c}\ddot{\theta}(\bbox{r},\bbox{r'})\nonumber \\
&+& \dot{\theta}(\bbox{r},\bbox{r'})
+\sin[\theta(\bbox{r},\bbox{r'})
-2\pi A(\bbox{r},\bbox{r'})],\label{RSJ}
\end{eqnarray}
plus Kirchhoff's current conservation conditions at each site.
Here the dots represent time derivatives. The three contributions
to  $i(\bbox{r}, \bbox{r'})$ are the displacement, the dissipative and
the superconducting currents, respectively. The phase difference across 
a junction is $\theta(\bbox{r},\bbox{r'})\equiv
\theta(\bbox{r})-\theta(\bbox{r'})$.
The currents are expressed in units of the junction critical current $I_{c}$;
time is measured in units of the characteristic time
$1/\omega_{c}=\hbar/(2eR_{n}I_{c})$, and
$\beta_{c}=(\omega_{c}/\omega_{p})^2$ is the Stewart-McCumber
parameter \cite{StMc}, with the plasma frequency $\omega_{p}$ defined as
$\omega_{p}^2=2eI_{c}/\hbar C$,  and $R_n$
is the junction's normal state resistance.  The bond frustration variable
$A(\bbox{r},\bbox{r'})$ is defined as the line integral of the vector potential $\bbox{A}$:
\begin{equation}
A(\bbox{r},\bbox{r'})=\frac{1}{\phi_{0}}
\int_{\bbox{r}}^{\bbox{r'}}\bbox{A}\cdot d\bbox{l},
\end{equation}
with the elementary quantum of flux $\phi_{0}=h/2e$. 
The frustration parameter $f$ measures the average flux piercing a
plaquette, measured in units of $\phi_{0}$.\\
The motivation for this paper is twofold: to study the dynamics of a
 few vortices in an array, and to see whether the results
 can be generalized  in order to explain  dynamical non-equilibrium 
states, like the axisymmetric coherent vortex state \cite{ACVS}.
 Here we deal mainly with the analysis
 of single-vortex voltage steps. The motion of a vortex
 produces a Faraday voltage across the array. 
In this paper we find three types of new steps. First we find 
single-vortex voltage steps. The voltage $V$ on these steps is
proportional to an integer multiple $n$ of the frequency of the ac drive:
\begin{equation}
V= \frac{n h \nu
}{2e}~~~~~~~~n=0,~1,~2, \cdot \cdot
\end{equation}
In the following we will also consider natural units $\hbar/2e=1$. Then
the voltage is normalized such that it corresponds to $2\pi$ times
the number of jumps between plaquettes of the vortex per
 time unit. This differs from another often used normalization by
 a factor $N_y$, the number of junctions perpendicular
 to the direction of the current injection. 
On such a step the motion of the single vortex is phase-locked to 
move an integer number of plaquettes per period $\frac{1}{\nu}$.
Next we find subharmonic single-vortex steps in the $I$-$V$'s 
of both overdamped and underdamped arrays. On these steps the voltage is:
 \begin{eqnarray}
V=\frac{n}{m} \frac{ h \nu}{2 e}~~~~~~~~~~~~&n&=0,~1,~2,~\cdot \cdot \\
\nonumber
&m&=1,~2,~\cdot \cdot
\end{eqnarray} 
The dynamics on these steps corresponds to the vortex 
moving $n$ plaquettes in $m$ periods.\\ 
Finally we simulate a vortex-antivortex pair and obtain the following steps: 
\begin{eqnarray} V=\frac{n N_y}{m}  \frac{h \nu}{2e} ~~~~~~~~&n&=0,~1,~2, \cdot \cdot \nonumber \\ 
 ~~~~~~~~~~~~~~~&m&=1,~2, \cdot \cdot
\end{eqnarray}\\ Each vortex moves $n N_y/2$ plaquettes every $m$ periods. \\
These new steps should be contrasted with other types 
of steps that arise in Josephson-junction arrays. 
In single junctions, harmonic ($V= n\frac{h \nu}{2e}$) voltage 
steps are present in both experimentally and numerically obtained
 $I$-$V$'s \cite{SHAP}. These steps are called single-junction
 Shapiro steps. On these steps the single junction has $n$ phase slips 
per period $\frac{1}{\nu}$.  In simulations of and experiments
 on ac+dc driven arrays so-called giant Shapiro steps have
 been observed \cite{GSS1,GSS_VL,GSSUB}. On these steps the 
$N_x N_y$ individual junctions along the direction of the external
 current are all phase-locked on the same single junction Shapiro step, and one 
obtains
\[V=  N_x N_y\frac{n h \nu
}{2e}~~~~~~~~n=0,~1,~2, \cdot \cdot\] 
The harmonic and subharmonic single-vortex voltage steps are therefore 
small in comparison to the giant Shapiro steps. Subharmonic giant Shapiro steps 
are observed experimentally \cite{GSSUB} and in simulations 
\cite{ACVS,HERIND}. A particular example of
these half-integer steps is found in the axisymmetric 
coherent vortex states (ACVS). These stationary states 
correspond  to an oscillating pattern of vortex 
and antivortex streets, which arrange themselves at a well-defined
angle with respect to the current direction \cite{ACVS}.\\
In studies of and experiments on ac+dc driven arrays, with an average
rational flux $\frac{p}{q} \phi_0$ 
piercing through each plaquette, giant {\it fractional} 
Shapiro steps were observed \cite{GSS_VL,GFSIM,GFLAD}.
On these steps the voltage is: 
\[V= \frac{ N_x N_y}{q} \frac{n h \nu}{2e} ~~~~~~~~~n=0,~1,~2, ~\cdot\cdot
\] An explanation for such steps has been proposed and verified
 in simulations \cite{GSS_VL,GFSIM}: an $f=\frac{p}{q}$ array
 has $q$ degenerate ground states, consisting of vortex lattices,
 carried into each other by translations. In one period of 
the driving current the vortex lattice moves from one degenerate state
 to the next. After $q$ periods every vortex has moved 
through the whole array, generating the observed voltage. \\
This shows that there are two kinds of Shapiro steps 
in Josephson-junction arrays.
Those based on the coherent phase slips of all the individual
junctions in the array (Giant Shapiro steps) and those 
involving coherent oscillatory
vortex motions (giant fractional Shapiro steps and ACVS). \\
\begin{figure}[t]
\unitlength=0.1in
\begin{picture}(20,30)
\put(52,-10){
\includegraphics{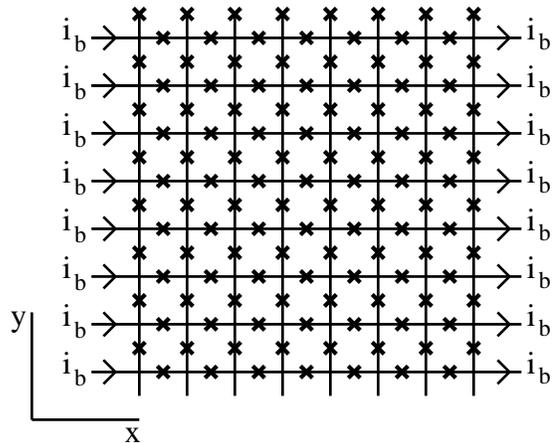}
}
\end{picture}
\caption{Array geometry used in the simulations,
 illustrated with an $8\times 8$ array. Junctions are 
denoted as crossed bonds. Periodic 
boundary conditions are imposed in the $y$-direction,
 while the current bias $i_b$ is applied along the $x$-direction. }
\label{fig1}
\end{figure}
The approach of our study is to have only one vortex, 
which makes it is possible
to separate the effect generic of vortex motion from the effects
of interaction between them. One can systematically study
the underlying microscopic dynamics of the phase-locked 
vortex motion. Then by considering a vortex-antivortex pair one 
can study the effect of interaction in its most simple form 
on the phase-locked states. 
The vortex dynamics on single-vortex voltage steps
has a number of new and interesting features,
e.g. we observe vortex motion with period doubling in underdamped arrays. 
The underlying microscopic vortex motion repeats itself only after $2$,
$4$, $8$, or even $16$ periods $\frac{1}{\nu}$,
although the vortex still moves, on average,
a fixed number of plaquettes per period.
In this paper we discuss how these steps can observed experimentally 
in arrays with a low density of vortices.\\
Previous authors \cite{DEPIN,SJMOD,Hag} have studied simple  
models of vortex motion. In these models the vortex is 
described as a point particle of mass $M(\beta_c)$ experiencing 
a certain friction, and moving in a sinusoidal potential.  
In Ref. \cite{Hag} we found that the $I$-$V$ of a dc driven array 
in the vortex regime is well described by such a 
vortex equation of motion in terms of the vortex-coordinate $y$ 
containing, instead of the usual linear viscous force, a 
nonlinear one 
\begin{equation}
M(\beta_c) \ddot{y} + \frac{A(\beta_c) \dot{y}}{1+B(\beta_c) \vert \dot{y}\vert } + i_0\sin 2\pi y = i 
\label{VJJ}
\end{equation}
In this equation $A(\beta_c)$ and $B(\beta_c)$ are phenomenological parameters.
This analysis is similar in spirit to the description of long Josephson junctions
in terms of the fluxon coordinates \cite{solit}.
 An interesting question previously left unaddressed, and considered in
this paper, is whether a phenomenological model for the vortex 
coordinate can still 
reproduce the $I$-$V$ of an ac+dc driven array. In this work we 
resolve this question and we compare the $I$-$V$ of an array with the
 result of Eq.$~$(\ref{VJJ}) with a time-dependent current 
$i=i_{dc}+i_{ac} \cos 2 \pi \nu t$, and using the parameters
 $M(\beta_c)$, $A(\beta_c)$ and $B(\beta_c)$ obtained for 
dc driven arrays. The results of Eq.$~$(\ref{VJJ}) and the simulations 
 are in reasonable agreement over a broad range  of values of the 
frequency and the amplitude of the ac drive. In other words 
the vortex experiences a nonlinear friction in an ac+dc driven
array. The outline of the paper is as follows.
In section II we discuss the calculational algorithm to compute 
the $I$-$V$. In section III we present the $I$-$V$'s containing the 
harmonic and subharmonic single-vortex steps and 
discuss the possibility of experimental verification.
 We then study subharmonic single-vortex voltage steps 
and period doubled vortex motion using 
the microscopic current distribution in the array as a function of time.
 Finally we compare the result of Eq.$~$(\ref{VJJ}) 
to those of the $I$-$V$'s obtained from the Josephson-junction array 
simulations. In section IV we investigate the effects of interaction 
and discuss the vortex-antivortex voltage steps in the $I$-$V$. In section 
V we present our conclusions.\\ 
\section{Calculational approach}
We numerically solve the equations of motion for a two-dimensional array.
We show the square lattice of
 $L_x \times L_y$ sites in Fig.$~$ \ref{fig1}. The sites are 
connected through Josephson junctions, denoted by crosses. We use the 
RCSJ-model of Eq.$~$(\ref{RSJ}) to relate the current  $i(\bbox{r},\bbox{r'})$
 through the Josephson junction to the phase difference 
$\theta(\bbox{r},\bbox{r'})$.
We use periodic boundary conditions (PBC) in the $y$-direction,
 while the current is fed in and taken out along the $x$-direction.
 This set of coupled nonlinear differential equations can be
 integrated efficiently using a fast Fourier transform algorithm 
\cite{ACVS,DickFFT}.
The array consists of $N_x\times N_y$ plaquettes ($N_x=L_x-1$ and $N_y=L_y$).
The vorticity $n(\bbox{ R})$ is defined as:
\begin{equation}
2\pi n(\bbox{R})=2\pi f +\sum_{{\cal P}(\bbox{R})}
\big(\theta(\bbox{r},\bbox{r'})-2\pi A(\bbox{r,r'})\big).
\label{vortdef}
\end{equation}
Here ${\cal P}(\bbox{R})$ denotes an anti-clockwise sum around 
the plaquette $\bbox{R}$ and the gauge-invariant phase difference
$\theta(\bbox{r},\bbox{r'})-2\pi A(\bbox{r},\bbox{r'})$ is
taken between $-\pi$ and $+\pi$.

\begin{figure}
\unitlength=0.1in
\begin{picture}(20,30)
\put(0,0){
\includegraphics{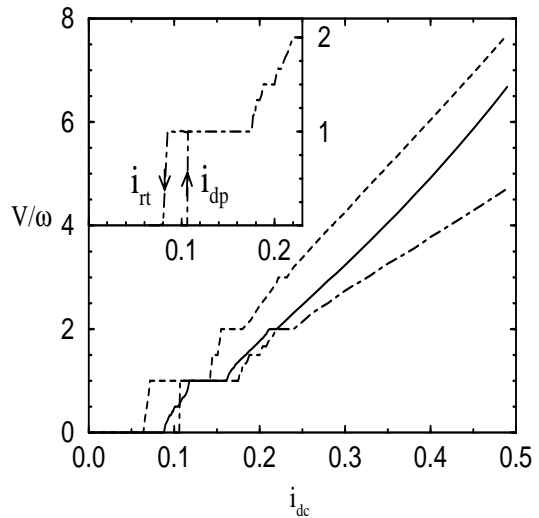}
}
\end{picture}
\caption{$I$-$V$'s obtained from simulation of a $16 \times 32$ 
array with parameters,  $i_{ac}=0.10$, $\nu=\frac{1}{25}$ and $\beta_c=0$ 
(continuous line); $\nu=\frac{1}{50}$ and $\beta_c=5$ (dotted line);
 and $\nu=\frac{1}{50}$ and $\beta_c=20$ (dashed-dotted line). 
The voltage in natural units is normalized by $\omega=2\pi \nu$. The inset contains 
a close-up of the $\beta_c=20$ curve,
 with an added $I$-$V$ branch showing hysteresis. }
\label{fig2}
\end{figure}
We are interested in the behavior of vortices in Josephson-junction
arrays. Stable vortices can be explicitly introduced in the initial
phase-configuration by the application of a small frustration \cite{Hag}.
The plaquette coordinate
$\bbox{R}$ with unit vorticity will be called the topological vortex coordinate.
The voltage $V(t)$ is obtained from:
\begin{equation}
V(t)=\sum_{y=0}^{L_y-1}\frac{d}{dt}[\theta (L_x-1,y)-\theta (0,y)]
\end{equation}
The time average of $V(t)$ is related to the average vortex velocity $v=\frac{V}{2\pi}$.  The microscopic dynamics of the vortex motion is reflected in the eddy-current distribution $C(\bbox{R},t)$,
\begin{equation}
C(\bbox{R},t)=\sum_{{\cal P}(\bbox{R})}i(\bbox{r,r'},t).
\end{equation}
Where ${\cal P}(\bbox{R})$ is the anti-clockwise sum over bonds about a dual lattice site $\bbox{R}$. The vortex shows up as a local extremum in the
eddy-current distribution.
\begin{figure}
\unitlength=0.1in
\begin{picture}(20,20)
\put(-3,0){
\includegraphics{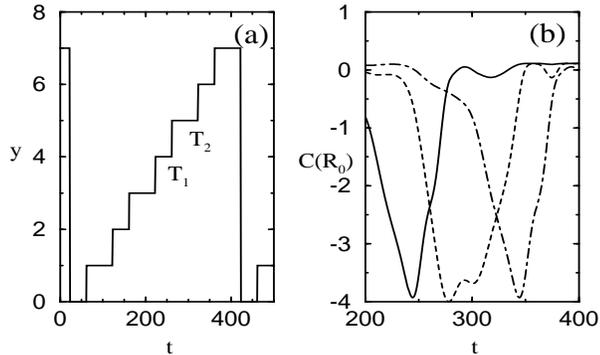}
}
\end{picture}
\caption{Period doubling on the $n=1$ step. 
The results are obtained from a simulation of an $8\times8$ array,
 with parameters $\beta_c=25$, $\nu=1/50$, $i_{ac}=0.10$, 
and $i_{dc}=0.13$. In (a) we plot the vortex position versus time. 
In (b) we plot the
 eddy currents versus time at three adjacent plaquettes of the middle column.
The continuous, dotted and dot-dashed lines are respectively 
 the first, second and third plaquette in the middle column of
 the array. The minima in these graphs indicate the position 
of the vortex.}
\label{fig10}
\end{figure}
\section{Single-vortex-induced voltage steps}
In this section we present the results of our simulations of a single vortex
in a Josephson-junction array.
 In the first subsection we discuss results for the $I$-$V$ characteristics.
  In the next subsection a detailed description of the single-vortex 
  voltage steps is given. We discuss the microscopic vortex motion,
 including its period doubling and the subharmonic single-vortex
 voltage steps. This is followed by a comparison of a simple
 model for vortex motion to the simulation results, and finally
 the discussion of possible experimental verification.
\subsection{The $I$-$V$'s}
The $I$-$V$'s are generated by gradually increasing
 the bias current $i_{dc}$ from $0$ starting with an initial configuration
containing a vortex.
We express the voltage in natural units
$\frac{\hbar}{2e}\equiv1$ and normalized by the frequency $2\pi
\nu$. The harmonic steps then occur at integer voltage values.\\
 For currents below the depinning current $i_{dp}$ the voltage is zero
(see Fig.$~$\ref{fig2}), 
when it is averaged over enough periods of the ac drive. 
The vortex deforms in response to the ac+dc drive, but stays
 in the same plaquette \cite{Dick}. Or, for low enough frequency 
$\nu$ and large enough $i_{ac}$,
it can even oscillate back and forth over a finite 
number of plaquettes.\\
The second branch of the $I$-$V$ is generated by decreasing the bias current.
As the initial phase configuration one uses the final phase
configuration obtained in the upward sweep.
 At the retrapping current $i_{rt}$ the average voltage returns to zero.
 In underdamped arrays the retrapping current $i_{rt}$ can be 
different from $i_{dp}$. This hysteretic behavior has often 
been  interpreted as evidence for inertia of the vortex in the dc 
driven case. 
An example of hysteresis is shown in the inset of Fig.$~$\ref{fig2}.
Focusing the discussion on the upward current sweep we encounter
 the first plateau at $V= 2 \pi \nu$. This is the first single-vortex
 step.
 The vortex moves on average one plaquette per period.
 There are two  more steps visible at multiples of $2 \pi \nu$ in 
Fig.$~$\ref{fig2}. Above $i_{dc}\approx 0.25$ no steps are visible 
anymore. The step width has become smaller than the current grid
size $\Delta i_{dc}=0.01$. 
The $\beta_c=0$ $I$-$V$ exhibits a pronounced upward curvature. Hence the
vortex viscosity is nonlinear \cite{Hag}.
Between the $n=1$ and $n=2$ step a small subharmonic step, 
at $V/2\pi \nu=\frac{3}{2}$, is visible in Fig.$~$\ref{fig2}.
On this step the vortex moves three plaquettes every two periods.
Using a smaller $\Delta i_{dc}=0.002$ one can even observe the $\frac{4}{3}$ and $\frac{5}{3}$ steps.\\
Increasing the damping parameter $\beta_c$ for given $\nu$, shifts
 the steps to higher values of the current bias $i_{dc}$. 
 This is due to an increase in the vortex viscosity with $\beta_c$ \cite{Hag}.
When $\beta_c$ is changed from $5$ to $20$, the $n=3$ 
step width is gradually reduced to below the current 
grid size $\Delta i_{dc}=0.01$.
For $\nu>1.0$ no integer steps are present
any more.
For these frequencies the velocity for which the vortex would
phase lock on the first integer step lies above the maximum vortex velocity
in the array.\\ 
 The width of a particular step varies in an oscillatory 
fashion as a function of $\frac{i_{ac}}{2 \pi \nu}$. This is
qualitatively
similar to the step width behavior of a single junction, which varies
as a Bessel function of $i_{ac}$.
By adjusting $i_{ac}$ one can make more steps visible.
\subsection{ Single-vortex-induced voltage steps}
\begin{figure}
\unitlength=0.1in
\begin{picture}(20,30)
\put(0,0){
\includegraphics{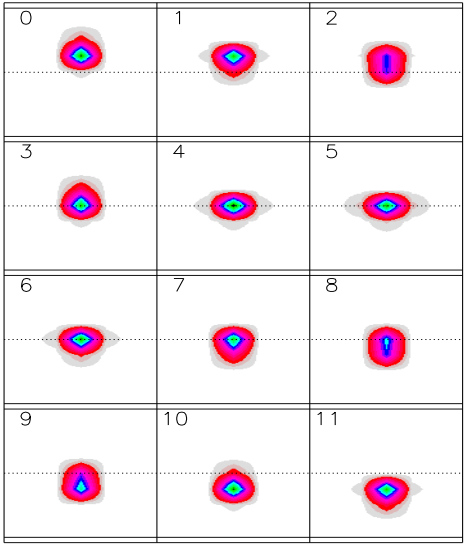}
}
\end{picture}
\caption{Period doubling on the $n=1$ step. Parameters are the same as
in Fig.$~$$3$.
 We show snapshots of the eddy current distributions at 
different times. The time step between two snapshots is 
$10$ (in units of $1/\omega_c$). The eddy-current distributions
are smoothed out by interpolation. }
\label{fig10ed}
\end{figure}

We now turn to the microscopic vortex motion on the single-vortex
voltage steps.
 We first consider the $n=0$ step, i.e. the response of a vortex
 that is pinned in one plaquette.
  One may expect the response to have the same period as the ac drive. 
In that case the quantity $v_p$, the voltage averaged over one period of the drive,
 is constant and equal to zero. We find, however, that $v_p$ can be
 non-zero on the $n=0$ step. This is repeated periodically in time.
 In table \ref{tab1} we show the periodicity of $v_p$,
 (in units of the driving period) 
 as a function of $i_{dc}$ for an $8\times 8$ array with parameters $\beta_c=25$, $\nu=1/50$, $i_{ac}=0.1$.\\ 
Next we focus on the $n=1$ step in table \ref{tab1}.
 The vortex depins at $i_{dc}\approx 0.12$. 
For $i_{dc}=0.13$ the voltage $v_p$ alternates between 
two different values. Figs.$~$\ref{fig10} and \ref{fig10ed} show the 
corresponding vortex 
motion. Fig.$~$\ref{fig10}(a) shows the topological vortex 
coordinate, defined in Eq.$~$(\ref{vortdef}), versus time. 
It shows that the time the vortex spends at one plaquette 
alternates between two values $T_1$ and $T_2$. The sum $T_1+T_2$ is equal to 
$2/\nu$. In Fig.$~$\ref{fig10}(b) we plot the eddy currents of three adjacent 
plaquettes versus time. In Fig.$~$\ref{fig10ed} we show snapshots of the 
spatial distribution of eddy currents. 
Frames $0$ and $10$ are almost equivalent.
A jump occurs between frames $2$ and $3$ and
between frames $8$ and $9$. The difference between $T_1$ (frames $0$-$2$) and
$T_2$ (frames $3$-$8$) is clearly visible.
All the figures show that the actual
 vortex motion is periodic with twice the period of the driving force. \\
Subharmonic steps turn out to be a generic feature 
of the $I$-$V$'s of ac+dc driven arrays. They are observed for
 different values of $i_{ac}$, $\nu$, $\beta_c$ and system sizes.
 Many different voltage values are possible.
We studied a $\frac{1}{2}$ step in detail.
 At $i_{dc}=0.11$ we found an $n=\frac{1}{2}$ step
in a $\beta_c=0$, $\nu=1/25$, $8\times 32$ array.  
On this step the generic behavior is as follows.
Although the vortex deforms significantly during the first period,
 the topological vortex coordinate does not change.
 Only in the second period it jumps to the neighboring plaquette.
 We now describe a typical example. 

\begin{figure}
\unitlength=0.1in
\begin{picture}(20,20)
\put(-3,0){
\includegraphics{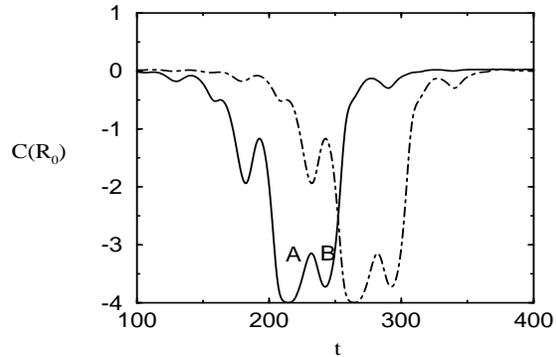}
}
\end{picture}
\caption{Vortex motion on a $\frac{1}{2}$ step for an $8\times 32$ array, 
with parameters $\beta_c=0$, $\nu=1/25$, $i_{ac}=0.10$, and 
$i_{dc}=0.11$.  We plot the 
eddy current versus time, as in Fig.$~$3(b), for 
two adjacent plaquettes (denoted by the continuous and the 
dot-dashed line respectively). The minima $A$ and $B$ in this graph 
indicate the position of the vortex. }
\label{fig14}
\end{figure}

In  Fig.$~$\ref{fig14} we show the eddy current versus time of the plaquette
 containing the vortex. Both curves have two minima: $A$ and $B$.
 In minimum $B$ the vortex has a different shape (and hence feels
 a different potential) than in $A$.
 From this we can deduce the following
 scenario. During the first period the vortex deforms from configuration 
 $A$ to
 $B$, staying in the same plaquette.
 In the following period it jumps to the next plaquette into the adjacent
 minimum $A$.
We also found a case ($i_{dc}=0.072$, $\nu=1/
50$) in which the vortex briefly
 jumped to the next plaquette (and returned to the original 
plaquette) during the first period. \\

\subsection{Comparison to a phenomenological model of vortex motion}
\begin{figure}
\unitlength=0.1in
\begin{picture}(20,25)
\put(-1,2){
\includegraphics{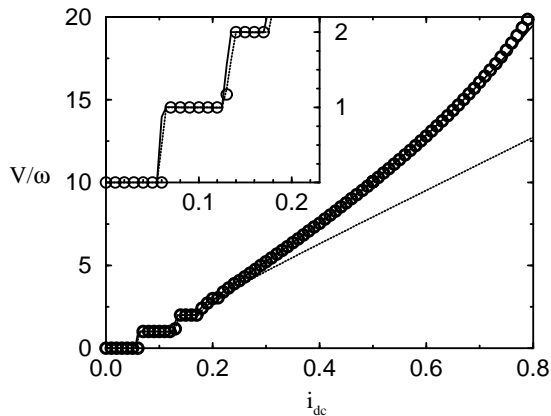}
}
\end{picture}
\caption{Comparison of results the model Eq.$~$(12) 
(dotted line, parameters: $\Omega=\pi/5$, $\beta=0.4$, and $\bar{i}_{ac}=1.0$)
 and (full line, parameters:
 $\Omega=0.617$, ${\cal B}=0.0424$, $\beta=0$, and $\bar{i}_{ac}=1.0$)
to the simulations of the full $16 \times 32$ array (circles, 
$\nu=\frac{1}{50}$, $\beta_c=2$, and $i_{ac}=0.10$). }
\label{fig18}
\end{figure}

In this subsection we compare the results of the numerical 
simulations to the results of a simple model for the vortex
 motion. The motion of a single vortex in an array can be modeled
 by that of a point particle with mass $M$ that, driven by a
 (Lorentz) force $i$, moves through a sinusoidal pinning 
potential and experiences a linear viscous damping force  
with constant viscosity coefficient $\eta$ \cite{DEPIN,SJMOD}.
 The vortex mass $M$ can be calculated by equating the 
electromagnetic energy  stored in the array to a vortex 
kinetic energy $\frac{1}{2} M {\dot y}^2$. The friction can 
be determined by equating the dissipated energy 
to $\eta {\dot y}^2$. This results in the following 
equation of motion for position $y(t)$, generalized to include
 a time dependent driving force $i$.
\begin{equation}
\pi \beta_c {\ddot y} + \pi {\dot y} + 
i_0\sin 2\pi y = i_{dc}+i_{ac} \cos 2 \pi \nu t.
\label{RJJ}
\end{equation}
For a square array $i_0\approx 0.10$ \cite{DEPIN}.
Every time the particle moves one plaquette ($y\rightarrow y+1$)
 an integrated $2\pi$ voltage pulse is generated across 
the array in the $x$-direction. The average of  $2\pi \dot{y}$
 is then the dimensionless average voltage measured across an array.\\
 We recently introduced a modified vortex equation 
of motion \cite{Hag} for dc driven vortices, including a nonlinear viscosity
\begin{equation}
M(\beta_c) \ddot{y} + \frac{A(\beta_c) \dot{y}}{1+B(\beta_c)\vert 
\dot{y}\vert } + i_0\sin 2\pi y = i_{dc}+i_{ac} \cos 2 \pi \nu t.
\label{VJJ2}
\end{equation}
The parameter values for overdamped case 
are $M(\beta_c)=0$, $A\approx 2.7$, $B\approx 1.8$, $i_0=0.1$ \cite{Hag}.\\ 
Equations (\ref{RJJ}) and (\ref{VJJ2}) are similar to 
the equations describing single Josephson junctions. To connect
 to the single-Josephson-junction literature, replace $y$ by
 a phase $\theta=2\pi y$, divide by $i_0$ and absorb the 
coefficient in front of the first derivative of $\theta$ in a new time unit:
\begin{equation}
{\beta} \ddot{\theta}\ + \frac{\dot{\theta}}{1+{\cal B}\dot{\theta}} + 
\sin \theta = {\bar i}_{dc}+{\bar i}_{ac} \cos \Omega \tau .
\label{VVJJ}
\end{equation}
Where $ \beta= M(\beta_c ) \frac{i_0 2\pi}{A^2}$, 
${\cal B}=\frac{ i_0 }{A} B$, $\Omega= \nu \frac{A }{i_0}$,
${\bar i}_{ac}=\frac{i_{ac}}{i_0}$ and ${\bar i}_{dc}=\frac{i_{dc}}{i_0}$.\\
To allow for a more quantitative comparison we use the procedure
described in Ref. \cite{Hag} to find the parameters $A$ and $B$ in Eq.$~$(\ref{VJJ2}).
 That is we fit the $I$-$V$ curve in the dc driven case to the form
 \[\frac{\sqrt{i_{dc}^2-i_0^2}}{A-B\sqrt{i_{dc}^2-i_0^2}}.\]
 Given these parameters one can numerically calculate the
 $I$-$V$ predicted by Eq.$~$(\ref{VJJ2}). In Fig.$~$\ref{fig18} 
we compare  these results for one specific $i_{ac}$, $\nu$ and $\beta_c$
 to the $I$-$V$ obtained from simulations of the full array.
 The results of Eq.$~$ (\ref{VJJ2}) are in good agreement with 
the simulation for larger currents $i_{dc}>0.3$. For this
 current regime the steps are too small with respect to 
the current grid $\Delta i_{dc}=0.01$ to be visible. \\
For lower currents the agreement is less satisfactory. 
The steps in the simulations and in Eq.$~$(\ref{VJJ2}) do overlap to a large extent,
but the lower edge of the steps 
is underestimated by  Eq.$~$(\ref{VJJ2}), especially at depinning.
However one finds large deviations when comparing the simulation data to 
Eq.$~$(\ref{RJJ}), the model with linear viscosity.\\
The phenomenological vortex mass for $\beta_c<35$ is 
$M(\beta_c)=0$, as found from  the dc $I$-$V$ curves in Ref. \cite{Hag}.
In the dc+ac case, however, we find small hysteresis loops in the $I$-$V$.
 One could interpret this as an indication for the presence
 of a non-zero $M(\beta_c)$.
 The hysteretic behavior of Eq.$~$(\ref{VJJ2}) is complicated.
 Using different $M\neq 0$ with the computed  values of 
$A(\beta_c) $ and  $B(\beta_c)$ did not yield better agreement
 with the simulation results.
 \subsection{Experimental verification}
In order to see how the single vortex phase-locking mechanism would
manifest itself in experimentally 
 accessible conditions, we now reinstate physical dimensions using 
experimental parameters \cite{EXP}.
Typical values for the parameters are $I_c=0.01-2.0 ~\mu A$,
 $I_c R_n= 300 ~\mu V$, $\omega_{c}=\frac{2eR_{n}I_{c}}{\hbar}\sim 1 $
 GHz
 and $\beta_c=10$-$100$.
The single-vortex voltage steps would occur for $i_{dc}=0.10-0.30 ~I_c$ and for frequencies 
$\nu< 0.10\cdot\omega_c \sim 100$ Mhz, and $\beta_c<50$.\\
Experiments are conducted at a finite temperature. 
Voltage steps are observed as a reduction
in the differential resistance 
$\frac{dV}{dI}$, while at zero temperature $\frac{dV}{dI}=0$ on
a step. 
 This reduction should be large enough, and extend over a sufficiently
 large current range, in order to be measurable. The temperature is 
expressed in units of $T_0=\frac{\hbar I_c}{2 e k_B}\approx 2\times 10^7 ~I_c$.
 We have performed a $T\ne 0$ simulation for one particular case:
 a $16 \times 16$ array with parameters $\nu=\frac{1}{25}$, 
$\beta_c=0$, and $i_{ac}=0.10$. The simulation was done using
 the algorithm introduced in Ref. \cite{HG}, as extended for arrays
\cite{ACVS}.
 We found that up to $T=0.004 ~T_0$ the $n=1$ step was clearly visible.
 For $T=0.008~T_0$ a reduction in $\frac{dV}{dI}$ was hardly discernible. 
 This puts an upper bound on the
appropriate temperatures, varying from $T=1$ to $200$ mK with the
value of $I_c$.\\
To study the single-vortex phase-locking phenomena experimentally,
one can use relatively small samples, 
to make sure that only a small absolute number of vortices is present
at any time, or a large array with a low vortex density.
In the former type of experiments the interaction
of the vortex with the boundaries, or
equivalently with image antivortices,
has to be taken into account.
When moving to a boundary, a vortex  may either escape from the array
or reflect as an antivortex \cite{bounce}.
In the parameter regime studied in this paper, vortices
are entering the array and leaving it again at the opposite boundary.
In practice the small number of vortices present 
will therefore fluctuate slightly.
This may effect the way in which phase-locking 
is established as compared to our simulation in which only one vortex
is present at any time.
To estimate this effect we have simulated a finite array in a small magnetic field 
and also found steps in the I-V's due to phase-locking with the
ac drive at 
voltages slightly different from those mentioned above.\\
 We also have performed simulations of 
triangular arrays, often used experimentally, and find qualitatively the 
same scenario for the occurrence of single-vortex voltage steps, 
as reported in section IIIA.\\
\begin{figure}
\unitlength=0.1in
\begin{picture}(20,25)
\put(-1,2){
\includegraphics{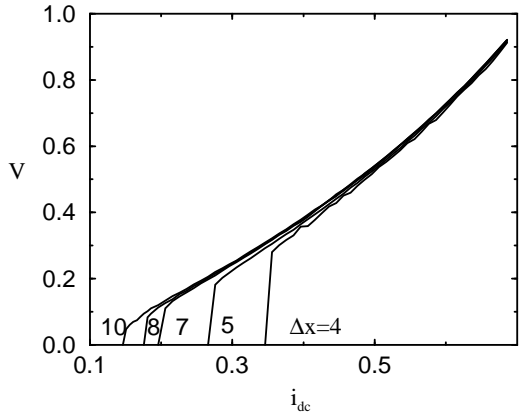}
}
\end{picture}
\caption{$I$-$V$ of a vortex and an antivortex separated by a 
distance $\Delta x=4$, $5$, $7$, $8$, and $10$ in an overdamped
 dc driven $32\times 32$ array. 
The annihilation current is the 
current at which the average voltage drops to zero in the $I$-$V$ of
the downward current
 sweep. }
\label{VAP2}
\end{figure}
\section{Vortex-antivortex pair-induced steps}
In this section we consider the effect of vortex-antivortex
 interaction
on the single-vortex voltage steps. In the first subsection 
we show the fractional voltage steps in the $I$-$V$ due to 
 the vortex-antivortex interaction, and obtain
 these steps from a phenomenological model for vortex motion including 
the logarithmic interaction. In the second subsection we briefly
 discuss the analogous behavior of excess and missing vortices 
in a checkerboard groundstate of an $f=\frac{1}{2}$ array. 

\subsection{Vortex-antivortex pair in an $f=0$ array}  

\begin{figure}
\unitlength=0.1in
\begin{picture}(20,25)
\put(-1,2){
\includegraphics{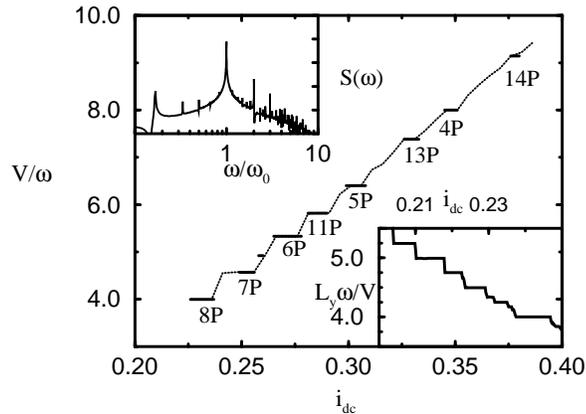}
}
\end{picture}
\caption{$I$-$V$ of a vortex-antivortex pair
with separation distance $\Delta x=7$ in
 an overdamped $32 \times 32$ array. The ac component of the 
driving current has frequency $\nu=\frac{1}{25}$ and amplitude
 $i_{ac}=0.10$. The labels mP  signify the underlying periodicity
 of the motion on these steps (i.e. the $m$ in Eq.$~$(5)).
In the upper inset we show the spectral
function of the voltage for the 6P step. 
In the lower inset we show the $I$-$V$ obtained from the model for
vortex-antivortex pair
 motion in the text. We used the following parameters:
 $L_y=32$, $2 \pi \nu=1.186$, ${\cal B}=0.0674$, and $\Delta x=8$. 
Here we plot, instead of $V$, $L_x \omega/V$. Steps 
then arise at the values $\frac{m}{n}$. The higher $n$ the 
smaller the step width.}
\label{VAP5}
\end{figure}

We calculate the zero temperature $I$-$V$'s 
of an array containing a vortex-antivortex pair.  There is 
no applied magnetic field, and the pair is included in the initial 
configuration by construction \cite{Hag}. In order to show how one 
may obtain meaningful results starting from such a metastable 
configuration, let us first discuss the case of only a dc drive. 
When the applied current is zero, the only mechanism that may prevent the
annihilation of the pair is the pinning of the lattice.
Pinning will prevent annihilation if the  mutual separation is at 
least 8 lattice constants. This minimum distance
is in approximate agreement with the value of
0.10 \cite{DEPIN} (in units of the junction critical current)
for the maximum pinning force on a vortex in an infinite system.
However, when the current is non-zero, and the vortices move
in different rows (in opposite directions), annihilation may
be absent even when the perpendicular distance between the
rows is less than 8 lattice constants, depending
on the magnitude of the current.
We have studied this stabilization by the driving force quantitatively
by 
performing downward current sweeps starting  from a number of initial 
configurations
containing pairs with different separation distances $\Delta x$. The
$I$-$V$'s are shown in Fig.$~$\ref{VAP2}. When the pair is unstable, and
thus annihilated, the average voltage obtained from the simulations
sharply drops to zero. This annihilation current strongly depends
on the separation distance of the pair in the initial configuration.
A higher current is
needed to stabilize a smaller pair. \\
Next we consider the behavior of an ac+dc driven pair. 
Since two vortices move
 in opposite directions under the influence of the current, 
their mutual distance will change in time.
Phase-locking can therefore not be established for a
voltage corresponding to one jump per period, since in the next
 period the environment has changed, and thereby the interaction
 strength. We nevertheless find steps
 in the $I$-$V$ of these systems.
 Once the
 vortex and the antivortex have traversed $L_y/2$ plaquettes,
 they cross each other again because we have periodic 
boundary conditions. 
Phase-locking is established if at that point the phases are the 
same again.
If phase-lock 
is established over $n$ traversals, and the motion is 
periodic with period $m$ (in units of the external frequency
 $\nu$), then the observed voltage is:
$V=n L_y/m 2 \pi \nu$. We can find each of these steps in 
the I-V in Fig.$~$\ref{VAP5}, although for higher $n$ the 
step width decreases rapidly. The step width is typically 
small ($\sim 0.01$).\\
When the vortex and antivortex are separated by a large distance,
one would expect to see single-vortex voltage steps again.
This does not occur for the system
size we considered due to the long range of the interaction.
In section IIIB we considered models for a single  vortex,
here we  consider a model for a pair. 
We model the vortex-antivortex interaction as being logarithmic. In addition 
we have to take into account the periodic boundary condition
along the $y$-direction: the vortex can see an antivortex 
behind it and in front of it (and vice versa). We denote the 
position of the vortex by $y_1$, and that of the antivortex 
by $y_2$. Let the coordinates take the values $0\le y_i<L_y$.
The images are then at $y_i\pm L_y$. We only take the direct
interaction and the interaction of the nearest image charge 
into account (i.e. at a distance less then $L_y$). The only 
stable current-driven pairs in the simulation are the ones
that move at a fixed separation $\Delta x$,  we therefore 
fixed the separation $\Delta x$ in this model. 
\newpage\noindent This leads to
the following equation of motion:
\begin{eqnarray}
\eta(\dot{y}_1) \dot{y}_1 &=  +i(t)-i_d \sin(2 \pi y_1) &-\frac{y_1-y_2}{{\Delta x}^2+(y_1-y_2)^2} \nonumber \\
&  &+ \epsilon \frac{L_y-\vert y_1-y_2 \vert}{{\Delta x}^2+(L_y-\vert y_1-y_2\vert)^2} \nonumber\\
\eta(\dot{y}_2) \dot{y}_2 &= -i(t)-i_d \sin(2 \pi y_2) &+\frac{y_1-y_2}{{\Delta x}^2+(y_1-y_2)^2} +  \nonumber\\
&  &-  \epsilon \frac{L_y-\vert y_1-y_2 \vert}{{\Delta x}^2+(L_y-\vert y_1-y_2\vert)^2} \nonumber
\end{eqnarray}
Here $\epsilon$ is the sign of $y_1-y_2$. We have computed the $I$-$V$'s
using these equations and we indeed found steps at  $V=n L_y/m 2 \pi \nu$. 
To capture the mechanism for the pair steps in the model, we found the essential 
ingredients to be the interaction and the restriction of
$y_i$ to $L_y$, that is periodic boundary conditions
in the $y$-direction.  The detailed form of the interaction 
determines the bias currents for which the steps occur.\\
\subsection{Defects in fully frustrated $f=\frac{1}{2}$ array  }
\begin{figure}
\unitlength=0.1in
\begin{picture}(20,25)
\put(-1,2){
\includegraphics{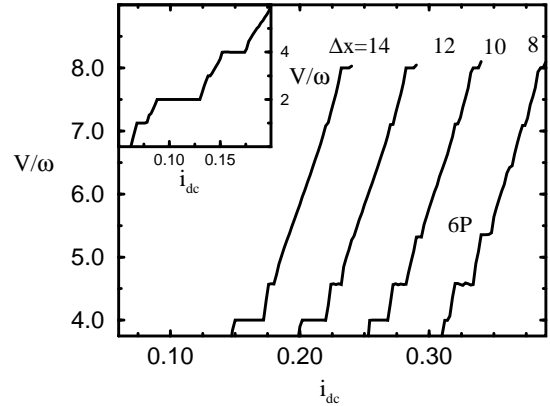}
}
\end{picture}
\caption{$I$-$V$ obtained by driving an excess and a missing
 vortex, separated by $\Delta x$,  in the vortex lattice of 
an overdamped fully frustrated $32 \times 32$ array. The ac 
component has frequency $\nu=\frac{1}{25}$ and amplitude 
$i_{ac}=0.10$. The curves are plotted for separations 
$\Delta x=8$, $10$, $12$, $14$. In the inset we show an
$I$-$V$ of an extra vortex moving in the groundstate
 vortex lattice of the fully frustrated  $16\times 16$ array. 
The frustration is $f=\frac{1}{2}+\frac{1}{15\times 16}$, 
$i_{ac}=0.10$, $\nu=\frac{1}{50}$, and $\beta_c=0$. }
\label{VAP4}
\end{figure}

It was found in \cite{V1SIM,Hag} that the $I$-$V$ of a single excess 
vortex in a dc driven $f=\frac{1}{2}$ array resembles that of 
single vortex in the $f=0$ case up to a current $i_{dc}=0.34$. At 
the latter current the entire checkerboard of vortices depins,
 producing a voltage that is much larger than
 the single vortex response.
The excess vortex moves 
in an even more nonlinear viscous fashion compared to $f=0$
\cite{Hag}.\\
Now we include an ac component to the dc driving current, and we find  
steps at $V= 2 n (2 \pi \nu)$. The factor two can be explained
 using the scenario proposed in \cite{GYORF}: first the excess
 vortex pushes the vortex in front of it to jump, and then follows 
suit. In total two jumps have been made. 
Small subharmonic $V= n (2 \pi \nu)$ ($n=1$,$3$) steps can be distinguished in the $I$-$V$ 
shown in the inset of Fig.$~$\ref{VAP4}.\\
We find that the $I$-$V$ of an excess vortex and a missing 
vortex (with respect to the checkerboard vortex lattice groundstate) 
in the $f=\frac{1}{2}$ case is similar to that of a vortex 
and an antivortex respectively in the $f=0$ case. 
An excess vortex can be annihilated by a missing vortex.
We calculated the $I$-$V$ for different 
values of $\Delta x$. In these $I$-$V$'s one again finds the
fractional vortex-antivortex-pair steps as in the $f=0$ case. One observes in Fig.$~$\ref{VAP4} that as 
the separation $\Delta x$ becomes larger, these 
steps disappear, and the steps corresponding to single vortices grow,
as one would intuitively expect. 
Finally only the latter remain. \\
\section{Summary and discussion}
In this paper we discussed the phase-locking  behavior of a 
single vortex and a vortex-antivortex-pair under the influence 
of an ac+dc drive. We obtained harmonic and subharmonic single-vortex
voltage steps, and studied the 
microscopic vortex dynamics. We found period doubled vortex motion. 
The $n=1$ step corresponds
to a vortex jumping one plaquette per period, since in a 
periodic array all the plaquettes in the same column along 
the periodic direction are equivalent, and one would expect 
period one behavior. The simulation, in which period doubling
 vortex motion is observed, was performed in an $8\times 8$ array
 with $\beta_c=25$. In this case the vortex motion can 
excite spin waves \cite{V1SIM,Uli}. These spin waves cause
 a spatially modulated environment for the vortex.
Similar phase-locking is observed in 
a ring of Josephson junctions \cite{Herre}.\\
 The spin waves may also be responsible for the complicated behavior
 of the hysteresis we observed in ac+dc driven arrays as a 
function of $\beta_c$.
In recent experiments subharmonic giant Shapiro steps \cite{GSSUB} were observed
in over- and underdamped arrays.
 In simulations it was shown that one can have 
such steps in a 2-D array by either including disorder 
\cite{ACVS} or inductive effects \cite{HERIND}, or generally
by any mechanism that breaks the translational invariance of 
the array. In overdamped single junctions  one can only 
get subharmonics by using 
non-sinusoidal current-phase relationships \cite{SJSUB}. 
Our simulations show that a single vortex itself may phase-lock
 on subharmonic steps. 
 The simulation of the pair case shows that single-vortex voltage steps
 are replaced by
 fractional steps, which can be understood in terms of vortex
 interaction. 
 Unfortunately this insight also makes it manifest
 why generalization to the dynamics associated 
 with Shapiro steps that involve many vortices
 (such as the axisymmetric coherent vortex states) is not feasible.
 Another consequence of the fact that we find subharmonic steps,
 is that a more realistic
 vortex equation of motion (than Eq.$~$(\ref{RJJ}) and (\ref{VJJ2}))
 should contain higher harmonic corrections to the sinusoidal
 potential. The presence of higher harmonics in the potential
 experienced by vortices was already noted by Lobb et al. 
\cite{DEPIN}.\\
We have shown that the parameters for which single-vortex voltage steps can
be observed are
 within the reach of present day experiments. 

\section*{Acknowledgements}
We thank D. Dom\'inguez, U. Geigenm\"uller, A. van Otterlo and A. van Oudenaarden for discussions.
This work was
supported in part by the Dutch organization for fundamental
research (FOM).
The work of JVJ has
been partially supported by NSF grant No. DMR-9521845. TJH
thanks the Bavarian ``FORSUPRA" program for financial support.

\begin{table}
\begin{tabular}{|l||l|l|} 
\hline
 $I_{dc}$ & step & period \\ 
 \hline \hline  
$0.0-0.06$ & $n=0$ & $1$ \\
$0.075$ & & $2$ \\
$0.08$ & & $4$ \\
$0.085$ & & $8$ \\ 
$0.088$ & & $16$ \\
$0.09$ & & $\sim 24$ \\
$0.095$ & & $8$ \\
$0.10$ & &$4$  \\
$0.105$,$0.11$ & & $2$ \\ 
$0.115$ & & $1$\\ \hline 
$0.125$ &$n=1$ & $4$ \\
$0.13-0.155$ & & $2$ \\
$0.165-0.19$ & & $1$ \\ \hline \hline
\end{tabular}
\vspace{0.2in}
\caption{ The periodicity of the voltage $v_p$ averaged over one period
of the driving force as a function of $i_{dc}$. The data is obtained from simulations of an $8\times8$ array, with parameters $\beta_c=25$, $\nu=1/50$, and $i_{ac}=0.10$.}
\label{tab1}
\end{table}


\begin{references}
\bibitem{EXP} These parameters are for
 arrays fabricated at Delft Technical University by 
A. van Oudenaarden. The bias current $I_{dc}$ can be measured
 with accuracy $\Delta I= 10^{-9} A$, the voltage $V$ 
with $\Delta V=10^{-7} V$. The feasible frequency range is $\nu=1$-$10 $MHz.
\end{references}
\end{document}